\documentclass[aps,prl,showpacs,twocolumn,superscriptaddress]{revtex4}  

\usepackage{graphicx}
\usepackage{dcolumn}
\usepackage{bm}
\usepackage{amssymb}   
\usepackage{epsfig}

\usepackage{url}
\newcommand{\fb} {{\rm fb}^{-1}}

\newcommand{\lumi} {6.8~\fb}

\newcommand{\prevlumi} {4.4~\fb}

\newcommand{\GeV}{{\rm GeV}}
\newcommand{\GeVc}{{\rm GeV/}c}

\newcommand{\MeVcsq}{{\rm MeV/}c^2}

\newcommand{\bz}{B^0}

\newcommand{\bp}{B^+}

\newcommand {\kst}{K^{*0}}
\newcommand {\kstp}{K^{*+}}
\newcommand {\kstall}{K^{*}}
\newcommand {\ks}{K_S^{0}}

\newcommand {\kstallpdg}{K^{*}(892)}

\newcommand{\kmm}{K^+ \mu^+ \mu^-}

\newcommand{\kstmm}{\kst \mu^+ \mu^-}
\newcommand{\kstpmm}{\kstp \mu^+ \mu^-}
\newcommand{\kstallmm}{\kstall \mu^+ \mu^-}

\newcommand{\smm}{s \mu \mu}
\newcommand{\bsmm}{b \to \smm}

\newcommand{\bkstll}{B \to K^{*}\ell\ell}

\newcommand{\bkstallmm}{B \to \kstallmm}
\newcommand{\bkmmincl}{B \to K^{(*)}\mu^+\mu^-}

\newcommand{\jpsi}{J/{\small \psi}}

\newcommand{\bpkmm}{\bp \to \kmm}

\newcommand{\bzkstmm}{\bz \to \kstmm}
\newcommand{\bpkstpmm}{\bp \to \kstpmm}

\newcommand{\fl}{F_{L}}
\newcommand{\afb}{A_{\mathit{FB}}}
\newcommand{\attwo}{A_{T}^{(2)} }
\newcommand{\aim}{A_{\mathit{im}}}

\newcommand{\cosm}{\cos \theta_\mu}
\newcommand{\cosk}{\cos \theta_K}
\newcommand{\cosmsq}{\cos^2 \theta_\mu}
\newcommand{\cosksq}{\cos^2 \theta_K}

\newcommand{\BaBar}{{\mbox{\sl B\hspace{-0.4em} {\small\sl A} \hspace{-0.37em}\sl B \hspace{-0.4em}}}{\small\sl A\hspace{-0.02em}R}}

\begin{document}

\title{\bf \boldmath
Measurements of the Angular Distributions in the Decays $\bkmmincl$ at CDF
}

\affiliation{Institute of Physics, Academia Sinica, Taipei, Taiwan 11529, Republic of China} 
\affiliation{Argonne National Laboratory, Argonne, Illinois 60439, USA} 
\affiliation{University of Athens, 157 71 Athens, Greece} 
\affiliation{Institut de Fisica d'Altes Energies, Universitat Autonoma de Barcelona, E-08193, Bellaterra (Barcelona), Spain} 
\affiliation{Baylor University, Waco, Texas 76798, USA} 
\affiliation{Istituto Nazionale di Fisica Nucleare Bologna, $^{bb}$University of Bologna, I-40127 Bologna, Italy} 
\affiliation{Brandeis University, Waltham, Massachusetts 02254, USA} 
\affiliation{University of California, Davis, Davis, California 95616, USA} 
\affiliation{University of California, Los Angeles, Los Angeles, California 90024, USA} 
\affiliation{Instituto de Fisica de Cantabria, CSIC-University of Cantabria, 39005 Santander, Spain} 
\affiliation{Carnegie Mellon University, Pittsburgh, Pennsylvania 15213, USA} 
\affiliation{Enrico Fermi Institute, University of Chicago, Chicago, Illinois 60637, USA}
\affiliation{Comenius University, 842 48 Bratislava, Slovakia; Institute of Experimental Physics, 040 01 Kosice, Slovakia} 
\affiliation{Joint Institute for Nuclear Research, RU-141980 Dubna, Russia} 
\affiliation{Duke University, Durham, North Carolina 27708, USA} 
\affiliation{Fermi National Accelerator Laboratory, Batavia, Illinois 60510, USA} 
\affiliation{University of Florida, Gainesville, Florida 32611, USA} 
\affiliation{Laboratori Nazionali di Frascati, Istituto Nazionale di Fisica Nucleare, I-00044 Frascati, Italy} 
\affiliation{University of Geneva, CH-1211 Geneva 4, Switzerland} 
\affiliation{Glasgow University, Glasgow G12 8QQ, United Kingdom} 
\affiliation{Harvard University, Cambridge, Massachusetts 02138, USA} 
\affiliation{Division of High Energy Physics, Department of Physics, University of Helsinki and Helsinki Institute of Physics, FIN-00014, Helsinki, Finland} 
\affiliation{University of Illinois, Urbana, Illinois 61801, USA} 
\affiliation{The Johns Hopkins University, Baltimore, Maryland 21218, USA} 
\affiliation{Institut f\"{u}r Experimentelle Kernphysik, Karlsruhe Institute of Technology, D-76131 Karlsruhe, Germany} 
\affiliation{Center for High Energy Physics: Kyungpook National University, Daegu 702-701, Korea; Seoul National University, Seoul 151-742, Korea; Sungkyunkwan University, Suwon 440-746, Korea; Korea Institute of Science and Technology Information, Daejeon 305-806, Korea; Chonnam National University, Gwangju 500-757, Korea; Chonbuk National University, Jeonju 561-756, Korea} 
\affiliation{Ernest Orlando Lawrence Berkeley National Laboratory, Berkeley, California 94720, USA} 
\affiliation{University of Liverpool, Liverpool L69 7ZE, United Kingdom} 
\affiliation{University College London, London WC1E 6BT, United Kingdom} 
\affiliation{Centro de Investigaciones Energeticas Medioambientales y Tecnologicas, E-28040 Madrid, Spain} 
\affiliation{Massachusetts Institute of Technology, Cambridge, Massachusetts 02139, USA} 
\affiliation{Institute of Particle Physics: McGill University, Montr\'{e}al, Qu\'{e}bec, Canada H3A~2T8; Simon Fraser University, Burnaby, British Columbia, Canada V5A~1S6; University of Toronto, Toronto, Ontario, Canada M5S~1A7; and TRIUMF, Vancouver, British Columbia, Canada V6T~2A3} 
\affiliation{University of Michigan, Ann Arbor, Michigan 48109, USA} 
\affiliation{Michigan State University, East Lansing, Michigan 48824, USA}
\affiliation{Institution for Theoretical and Experimental Physics, ITEP, Moscow 117259, Russia}
\affiliation{University of New Mexico, Albuquerque, New Mexico 87131, USA} 
\affiliation{Northwestern University, Evanston, Illinois 60208, USA} 
\affiliation{The Ohio State University, Columbus, Ohio 43210, USA} 
\affiliation{Okayama University, Okayama 700-8530, Japan} 
\affiliation{Osaka City University, Osaka 588, Japan} 
\affiliation{University of Oxford, Oxford OX1 3RH, United Kingdom} 
\affiliation{Istituto Nazionale di Fisica Nucleare, Sezione di Padova-Trento, $^{cc}$University of Padova, I-35131 Padova, Italy} 
\affiliation{LPNHE, Universite Pierre et Marie Curie/IN2P3-CNRS, UMR7585, Paris, F-75252 France} 
\affiliation{University of Pennsylvania, Philadelphia, Pennsylvania 19104, USA}
\affiliation{Istituto Nazionale di Fisica Nucleare Pisa, $^{dd}$University of Pisa, $^{ee}$University of Siena and $^{ff}$Scuola Normale Superiore, I-56127 Pisa, Italy} 
\affiliation{University of Pittsburgh, Pittsburgh, Pennsylvania 15260, USA} 
\affiliation{Purdue University, West Lafayette, Indiana 47907, USA} 
\affiliation{University of Rochester, Rochester, New York 14627, USA} 
\affiliation{The Rockefeller University, New York, New York 10065, USA} 
\affiliation{Istituto Nazionale di Fisica Nucleare, Sezione di Roma 1, $^{gg}$Sapienza Universit\`{a} di Roma, I-00185 Roma, Italy} 

\affiliation{Rutgers University, Piscataway, New Jersey 08855, USA} 
\affiliation{Texas A\&M University, College Station, Texas 77843, USA} 
\affiliation{Istituto Nazionale di Fisica Nucleare Trieste/Udine, I-34100 Trieste, $^{hh}$University of Trieste/Udine, I-33100 Udine, Italy} 
\affiliation{University of Tsukuba, Tsukuba, Ibaraki 305, Japan} 
\affiliation{Tufts University, Medford, Massachusetts 02155, USA} 
\affiliation{Waseda University, Tokyo 169, Japan} 
\affiliation{Wayne State University, Detroit, Michigan 48201, USA} 
\affiliation{University of Wisconsin, Madison, Wisconsin 53706, USA} 
\affiliation{Yale University, New Haven, Connecticut 06520, USA} 
\author{T.~Aaltonen}
\affiliation{Division of High Energy Physics, Department of Physics, University of Helsinki and Helsinki Institute of Physics, FIN-00014, Helsinki, Finland}
\author{B.~\'{A}lvarez~Gonz\'{a}lez$^v$}
\affiliation{Instituto de Fisica de Cantabria, CSIC-University of Cantabria, 39005 Santander, Spain}
\author{S.~Amerio}
\affiliation{Istituto Nazionale di Fisica Nucleare, Sezione di Padova-Trento, $^{cc}$University of Padova, I-35131 Padova, Italy} 

\author{D.~Amidei}
\affiliation{University of Michigan, Ann Arbor, Michigan 48109, USA}
\author{A.~Anastassov}
\affiliation{Northwestern University, Evanston, Illinois 60208, USA}
\author{A.~Annovi}
\affiliation{Laboratori Nazionali di Frascati, Istituto Nazionale di Fisica Nucleare, I-00044 Frascati, Italy}
\author{J.~Antos}
\affiliation{Comenius University, 842 48 Bratislava, Slovakia; Institute of Experimental Physics, 040 01 Kosice, Slovakia}
\author{G.~Apollinari}
\affiliation{Fermi National Accelerator Laboratory, Batavia, Illinois 60510, USA}
\author{J.A.~Appel}
\affiliation{Fermi National Accelerator Laboratory, Batavia, Illinois 60510, USA}
\author{A.~Apresyan}
\affiliation{Purdue University, West Lafayette, Indiana 47907, USA}
% ...
\collaboration{CDF Collaboration\footnote{With visitors from $^a$University of Massachusetts Amherst, Amherst, Massachusetts 01003,
$^b$Istituto Nazionale di Fisica Nucleare, Sezione di Cagliari, 09042 Monserrato (Cagliari), Italy,
$^c$University of California Irvine, Irvine, CA  92697, 
$^d$University of California Santa Barbara, Santa Barbara, CA 93106
$^e$University of California Santa Cruz, Santa Cruz, CA  95064,
$^f$CERN,CH-1211 Geneva, Switzerland,
$^g$Cornell University, Ithaca, NY  14853, 
$^h$University of Cyprus, Nicosia CY-1678, Cyprus, 
$^i$University College Dublin, Dublin 4, Ireland,
$^j$University of Fukui, Fukui City, Fukui Prefecture, Japan 910-0017,
$^k$Universidad Iberoamericana, Mexico D.F., Mexico,
$^l$Iowa State University, Ames, IA  50011,
$^m$University of Iowa, Iowa City, IA  52242,
$^n$Kinki University, Higashi-Osaka City, Japan 577-8502,
$^o$Kansas State University, Manhattan, KS 66506,
$^p$University of Manchester, Manchester M13 9PL, England,
$^q$Queen Mary, University of London, London, E1 4NS, England,
$^r$Muons, Inc., Batavia, IL 60510,
$^s$Nagasaki Institute of Applied Science, Nagasaki, Japan, 
$^t$National Research Nuclear University, Moscow, Russia,
$^u$University of Notre Dame, Notre Dame, IN 46556,
$^v$Universidad de Oviedo, E-33007 Oviedo, Spain, 
$^w$Texas Tech University, Lubbock, TX  79609, 
$^x$IFIC(CSIC-Universitat de Valencia), 56071 Valencia, Spain,
$^y$Universidad Tecnica Federico Santa Maria, 110v Valparaiso, Chile,
$^z$University of Virginia, Charlottesville, VA  22906,
$^{aa}$Yarmouk University, Irbid 211-63, Jordan,
$^{ii}$On leave from J.~Stefan Institute, Ljubljana, Slovenia, 
}}
\noaffiliation

\begin{abstract}
We reconstruct 
the decays $B \to K^{(*)} \mu^+ \mu^-$ and measure their angular distributions in $p\bar{p}$ collisions at $\sqrt{s}=1.96~{\rm TeV}$
using a  data sample corresponding to an integrated luminosity of $\lumi$.
The  transverse polarization asymmetry $\attwo$ and the time-reversal-odd charge-and-parity asymmetry $\aim$  
are measured for the first time, together with the $\kstall$ longitudinal polarization fraction $\fl$ and the
muon forward-backward asymmetry $\afb$, for the decays $\bzkstmm$ and $\bpkstpmm$.
Our results  are among the most accurate to date and consistent with those from other experiments.
\end{abstract}

\pacs{13.25 Hw, 13.20 He}
\maketitle

The decays $\bkmmincl$, which proceed via the flavor-changing neutral-current (FCNC) process $\bsmm$,
are considered among the most promising probes of the standard model (SM) and its extensions.
In the SM they occur through higher order amplitudes,
though additional processes with beyond-the-standard-model (BSM)
contributions could arise.
One can obtain sensitivity to BSM physics from precise measurement of the decay amplitudes,
especially the angular distributions of the decay products.

The full differential decay distribution for 
the decay 
$B \to \kstallpdg \mu^+ \mu^- \to K \pi \mu^+ \mu^-$
is described by four independent kinematic variables:
the dimuon invariant
mass  squared $q^2 \equiv M_{\mu\mu}^2c^2 (\GeV^2/c^2)$, 
the angle $\theta_\mu$  between the $\mu^+$ ($\mu^-$) direction 
and the direction opposite to the $B$ ($\overline{B}$) meson 
in the dimuon rest frame,  
the angle $\theta_K$  between the kaon direction and the
direction opposite to the $B$ meson in the $\kstall$ rest frame, 
and the angle $\phi$  
between the two planes formed by the dimuon and the $K$-$\pi$ systems.
The angle $\phi$ is zero if the two planes are parallel.
The 
distributions of $\theta_\mu$, $\theta_K$, and  $\phi$ are projected from the full differential
decay distribution and can be parametrized with four angular observables, 
$\afb$, $\fl$, $\attwo$, and  $\aim$~\cite{Kruger:2005ep_APS,Egede:2008uy_APS,DescotesGenon:2011yn_APS,Kruger:1999xa_APS,Altmannshofer:2008dz_APS}:
\begin{eqnarray}
\label{eq:afb}
\frac{1}{\Gamma}
\frac{d\Gamma}{ d\cosk} = \frac{3}{2} \, \fl \cosksq + \frac{3}{4} \, (1-\fl)(1-\cosksq) \nonumber, \\
\frac{1}{\Gamma}
\frac{d\Gamma}{ d\cosm} = \frac{3}{4} \, \fl(1-\cosmsq)+ \frac{3}{8} \, (1-\fl)(1+\cosmsq) \nonumber \\
+ \afb \cosm, \nonumber \\
\frac{1}{\Gamma}
\frac{d\Gamma}{ d\phi}  =\frac{1}{2\pi} \left[ 1+\frac{1}{2}(1-\fl)\attwo \cos2\phi + \aim \sin2\phi \right],
\label{eq:angular}
\end{eqnarray}

where $\Gamma \equiv \Gamma(\bkstallmm)$,  $\afb$ is the  muon
forward-backward asymmetry, 
$\fl$ is the $\kstall$ longitudinal polarization fraction,
$\attwo$ is the  transverse polarization asymmetry~\cite{Kruger:2005ep_APS},
and  $\aim$ 
is
the time-reversal-odd charge-and-parity asymmetry ($T$-odd $CP$ asymmetry)~\cite{Bobeth:2008ij_APS,Egede:2008uy_APS}.
In the case of $\bar{B}$ decay, $\sin\phi$ has opposite sign to that of the $B$ decay due to the flipped dimuon plane.
Since $\aim$ is a coefficient of the $\sin2\phi$ term in Eq.~(\ref{eq:afb}),
 $\aim$ is a measure of the $CP$ asymmetry.
Experimental access to these angular observables extends the
sensitivity to a large class of BSM physics models

The differential decay distribution is calculated in an operator product expansion~\cite{Wilson:1969zs}.
In the SM, the nonzero terms are parametrized by the short-distance Wilson coefficients $C^{\rm eff}_{7,9,10}$.
Since various BSM models predict couplings to the $\kst$ helicity states 
that are different from the SM ones,
the measurement of $\fl$ could constrain BSM parameters~\cite{Kruger:2005ep_APS}.
The SM prediction in the range $1 < q^2 < 6\,\GeV^2/c^2$ is $\fl^{\rm SM}=0.73^{+0.02}_{-0.03}$~\cite{DescotesGenon:2011yn_APS}.
The quantity $\afb$ is expected to be small at low $q^2$
and to have a large positive value at high $q^2$ in the SM.
In the range $1 < q^2 < 6\,\GeV^2/c^2$,
the SM predicts $\afb^{\rm SM}=0.022\pm0.028$~\cite{DescotesGenon:2011yn_APS}.
The BSM contributions can change the magnitude and the sign of $\afb$.
For example, some BSM models such as supergravity models with large $\tan\beta$~\cite{Ali:1999mm_APS}
allow an opposite sign of $C^{\rm eff}_7$ compared to the SM,
resulting  in the opposite sign of $\afb$ at low $q^2$.
In the SM $\attwo$ is expected to be approximately zero at low $q^2$ and negative at high $q^2$~\cite{Bobeth:2010wg_APS}.
Some BSM models like $R$-parity conserving minimal supersymmetry~\cite{Lunghi:2006hc_APS}
predict the existence of right-handed currents.
In such a case, $\attwo$ could be enhanced
 up to $\pm 1$~\cite{Kruger:2005ep_APS,Alok:2010zd_APS,DescotesGenon:2011yn_APS}.
The SM predicts the quantity $\aim$ to be very close to zero for
all accessible $q^2$, and this quantity is particularly sensitive to
the existence of $CP$ violation in right-handed currents in BSM physics~\cite{Bobeth:2008ij_APS,Altmannshofer:2008dz_APS}.

\BaBar~\cite{Aubert:2008ju_APS}, Belle~\cite{Wei:2009zv}, and CDF~\cite{Aaltonen:2011cn} have  reported measurements of 
$\afb$  and $\fl$
in the $\bkstll$ 
decay modes.
All experiments find $\afb$ to be larger
than the SM expectation, but so far none has sufficient sensitivity to be conclusive.
In this Letter, we report measurements of the
angular distributions in the decay $\bkmmincl$
using  a data sample, corresponding to an integrated $\lumi$, 
of $p\bar{p}$ collisions at a center-of-mass energy of $\sqrt{s}=1.96~{\rm TeV}$,
collected with the CDF~II detector between March 2002 and June 2010.
The measurement updates and supersedes an earlier
analysis on $\prevlumi$~\cite{Aaltonen:2011cn}.
Besides the 54\% increase in luminosity, we have added new
decays channels $\bpkstpmm$
and improved the selection achieving a 9\% increase in $\bzkstmm$ signal
 efficiency with almost the same background rejection.
The resulting factor of 82\% increase in
 signal statistics allows us to access
for the first time the angular observables
$\attwo$ and $\aim$ in the decays $\bkstallmm$.
The measurements of the total and differential branching ratios of the $\bkmmincl$ decays are reported in another letter~\cite{Aaltonen:2011qs_APS}.

The reconstruction of the $\bkmmincl$ decays starts with a
dimuon sample selected by the online trigger system~\cite{CDF:Trigger} of the CDF~II
detector~\cite{Acosta:2004yw_APS}.  
The trigger system uses information from muon detectors 
and the central drift tracking chamber.
The central drift tracking chamber~\cite{Affolder:2003ep_APS} provides 96 measurements per track between
radii of 40 and 137~cm, allowing an accurate determination of the charged-particle momentum.
The CMU and CMX muon drift chambers~\cite{Ascoli:1987av_Dorigo:2000ip_APS}
cover 
the pseudorapidity regions
$|\eta| < 0.6$ and $0.6 < |\eta| < 1.0$,  respectively~\cite{CDF_coordinate}.
The CMP muon system
is located radially behind the CMU and an additional steel absorber,
and covers $|\eta|<0.6$.
The dimuon trigger requires a pair of oppositely charged tracks
with momenta transverse to the beamline
$p_T\geq1.5\,\GeVc$, that are also identified in the CMU or CMX chambers.
At least one of the muon tracks in the pair is required to
be a CMU muon.  The trigger also requires that either the  dimuon pair satisfies
$L_{xy}>100~\mu \rm m$, where the transverse  decay length  $L_{xy}$ is 
the flight distance between the dimuon vertex and the event primary vertex
and projected on the dimuon momentum vector,
or that one of the muon candidates has $p_T>3.0\,\GeVc$ and  is identified by both CMU and CMP chambers.
The other detector subsystems relevant for this analysis are discussed in Ref.~\cite{Aaltonen:2009tz_ref}.
Each offline track is required to satisfy the standard quality requirements (e.g. hits in the silicon detector)
 to ensure well measured momenta and decay vertices~\cite{Aaltonen:2011cn}.
The decay length and invariant mass of each dimuon pair are calculated after a vertex fit
using the muon tracks. 
Dimuons are required to have $q^2$ values outside the ranges of  
$ 8.68 <q^2 < 10.09 \,\GeV^2/c^2$ and
$12.86 <q^2 < 14.18 \,\GeV^2/c^2$~\cite{Aaltonen:2011cn},
to be inconsistent with decays of
either $\jpsi$  or $\psi(2S)$ mesons,
which are typically reconstructed with $14\,\MeVcsq$ mass resolution.
The dimuon pair is then combined with charged tracks forming
a $\kst \to K^+\pi^-$ candidate to form a $\bzkstmm$ candidate, or
a $\kstp \to \ks(\to \pi^+\pi^-)\pi^+$ candidate to form a $\bpkstpmm$ candidate.
Charge-conjugate modes are implied throughout this Letter.
The  $\ks$, $\kst$, and $\kstp$ candidates 
are required to have reconstructed invariant masses consistent with 
the world average values~\cite{Nakamura:2010zzi_APS},
and to have $p_T>1\,\GeVc$. 
The $\ks$ is also required to have its
decay vertex displaced from the dimuon vertex. 
The ambiguity in the mass assignment of the decay products in the $\kst \to K^+\pi^-$ decay 
is handled by choosing the combination whose
$K^+\pi^-$ mass is closest to the world average $\kst$ mass, which is
correct in approximately 92\% of cases.
The reconstructed $B$ candidates are required to have $p_T>4\,\GeVc$. 
To further optimize the event selection,
an artificial neural network (NN)
classifier is trained for each channel
using simulated signal events and background events
taken from mass sidebands in data.
Simulated signal distributions are corrected using scale factors 
extracted by comparing simulation to data in the 
control samples $B \to \jpsi(\to \mu^+\mu^-) K^{(*)}$.
The optimized NN threshold is determined 
to minimize the statistical uncertainty of the angular observables,
using many kinematic observables including $p_T$, invariant mass, vertex fit
parameters, and muon identification quality~\cite{Aaltonen:2011cn}. 

\begin{figure}[tpb]
  \begin{center}
    \begin{tabular}{ll}
      \resizebox{.235\textwidth}{!}{\includegraphics[trim=0 0 0 0,scale=1.00,clip]{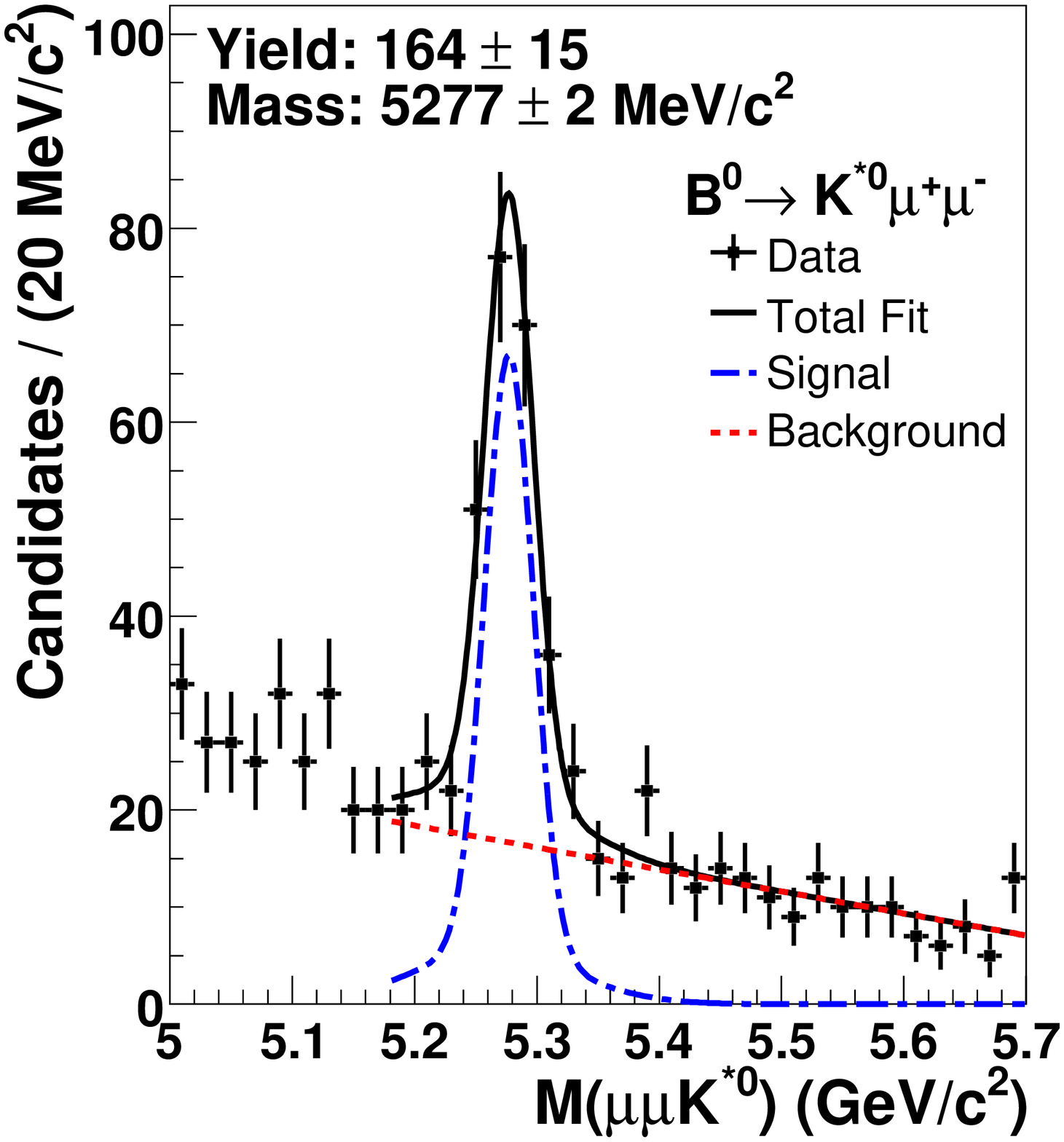}}& 
      \resizebox{.235\textwidth}{!}{\includegraphics[trim=0 0 0 0,scale=1.20,clip]{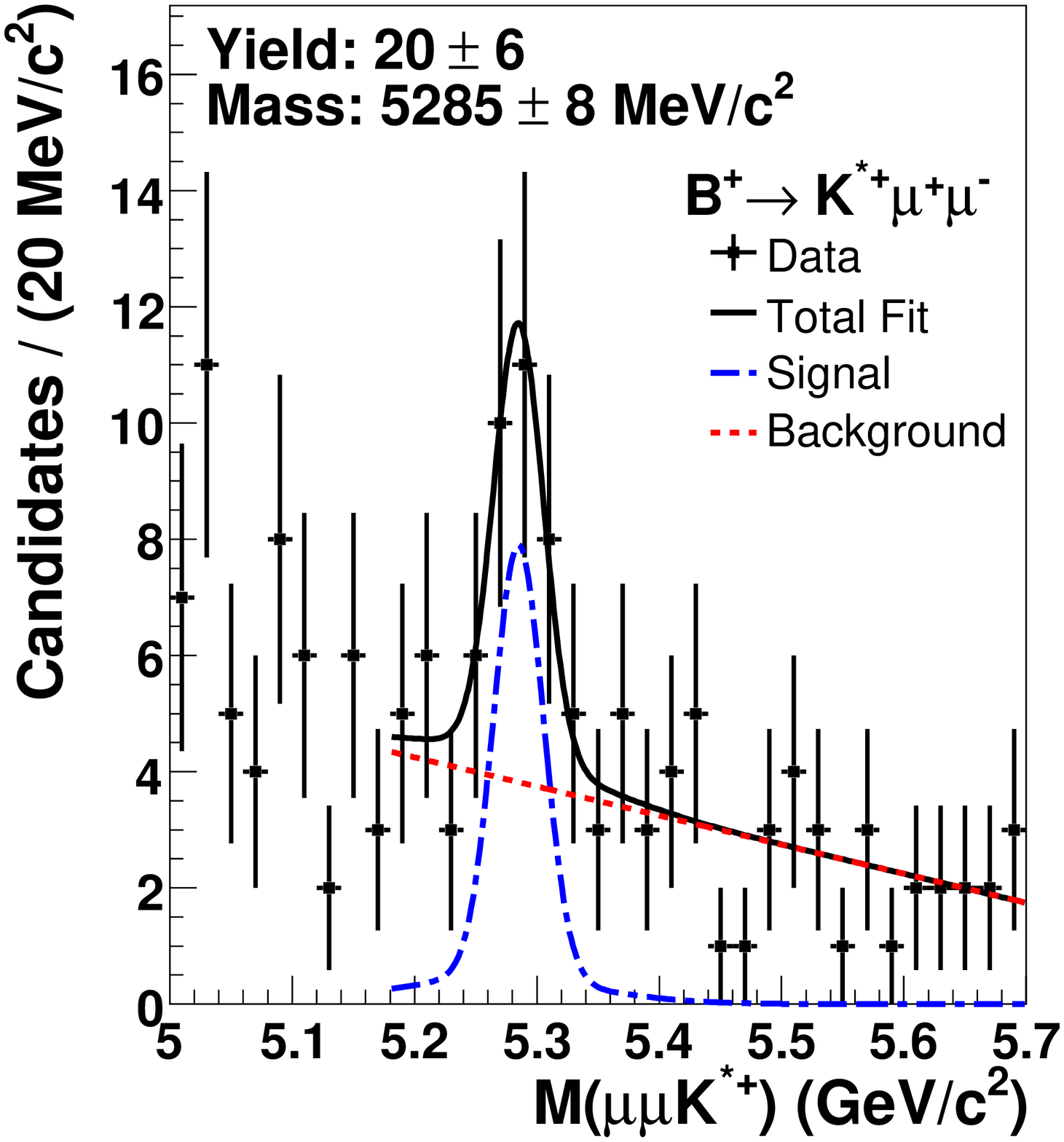}}
    \end{tabular}
  \end{center}
  \caption{
Invariant mass of $\bzkstmm$ (left) and $\bpkstpmm$ (right), with fit results overlaid.
}
  \label{fig:rarebmass_data}
\end{figure}

The signal yield is obtained by an unbinned maximum likelihood fit
to the invariant mass distribution with a probability density
function (PDF) consisting of Gaussian distributions for the signal 
and a linear background.
We find a total of $234\pm19$ $\bpkmm$,
$164\pm15$ $\bzkstmm$,  
and $20\pm 6$ $\bpkstpmm$ events.
We divide the signal region into six bins in $q^2$. 
Two semi-inclusive bins are included with ranges covering theoretically well-controlled regions.
We obtain the signal yields in the individual $q^2$ ranges 
by fitting the invariant mass in a similar way.
The resulting yields are listed in Tables~\ref{tab:summary_kstmm}, \ref{tab:summary_kstmm_all}, and \ref{tab:summary_kmm}.
The invariant mass distributions are shown in Fig.~\ref{fig:rarebmass_data}.  

To extract the quantities $\fl$, $\afb$, $\attwo$, and $\aim$,
we perform  likelihood fits to distributions of $\cosk$, $\cosm$, and $\phi$
for events in each $q^2$ range. 
The signal fractions are fixed to the values obtained from the invariant mass fits.
Signal PDFs for angular distributions are formed from Eq.~(\ref{eq:afb}), 
taking into account the estimation of the detector acceptance
of the decay angles and the $K$-$\pi$ interchange using Monte Carlo simulations.
The incorrect $K$-$\pi$ assignment in the $\kst \to K^+ \pi^-$ decay 
distorts the signal mass distribution and swaps the sign of $\cosm$.
This effect is taken into account by introducing an additional signal-like term
into the likelihood function. The contribution from decays with non-resonant $K$-$\pi$ pairs
is estimated to be small~\cite{Kruger:2005ep_APS} and neglected in the fit.
The  background PDF shapes for the angular distributions are modeled
from events in the $B$ mass sidebands.
The values of $\fl$ in individual
$q^2$ ranges are  extracted from  fits to the $\cosk$ distributions and then used as inputs 
in the determinations of the other three observables.  
 The asymmetry $\afb$ is obtained from fits to the $\cosm$ distributions, 
and $\attwo$ and $\aim$ are from fits to 
the $\phi$ distributions.

\begin{figure}[tpb]
  \begin{center}
\setlength{\tabcolsep}{1pt}
\begin{tabular}{c}
\includegraphics[scale=0.245,trim=8 5 57 48,clip]{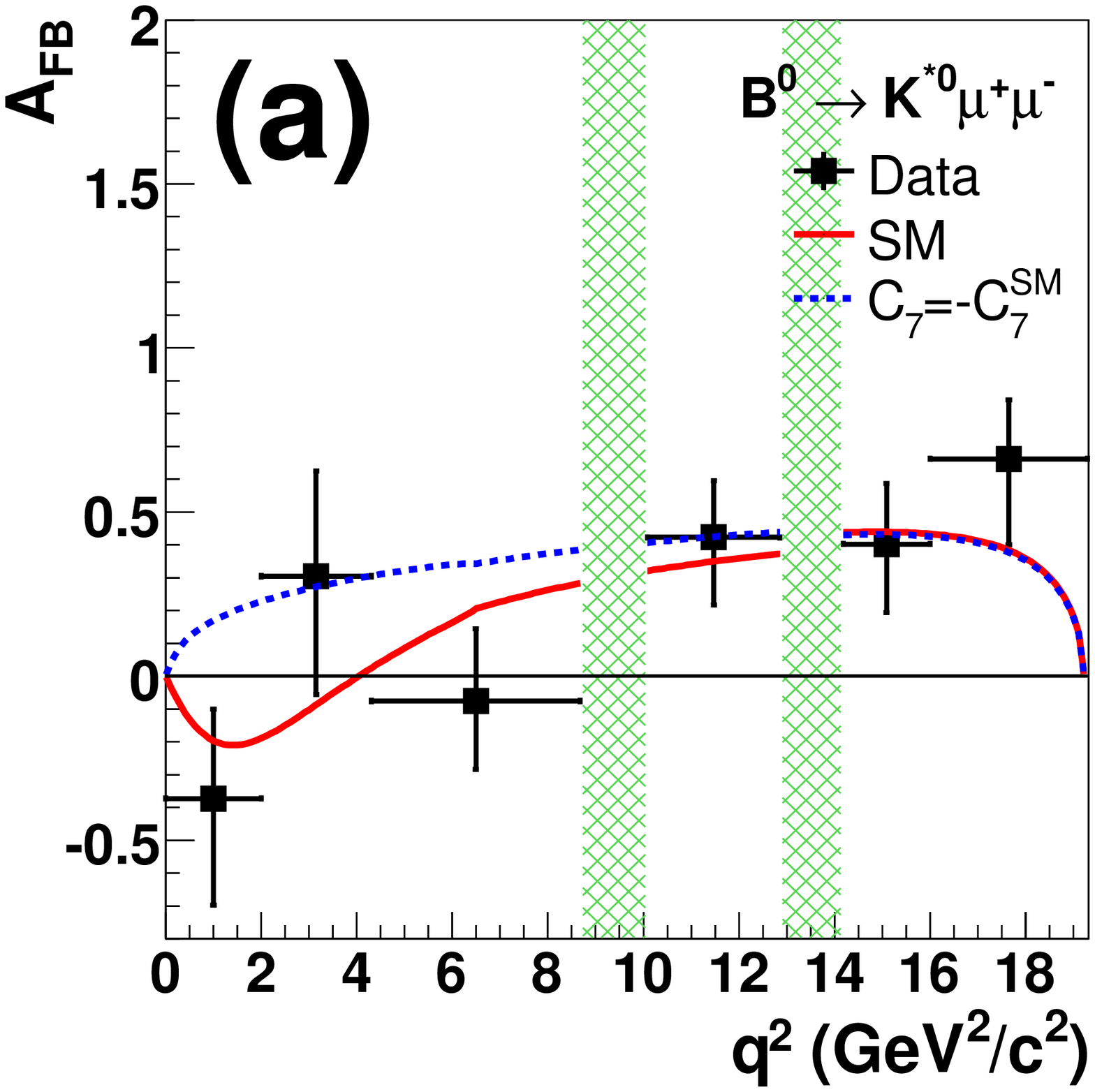}
\includegraphics[scale=0.245,trim=8 5 55 48,clip]{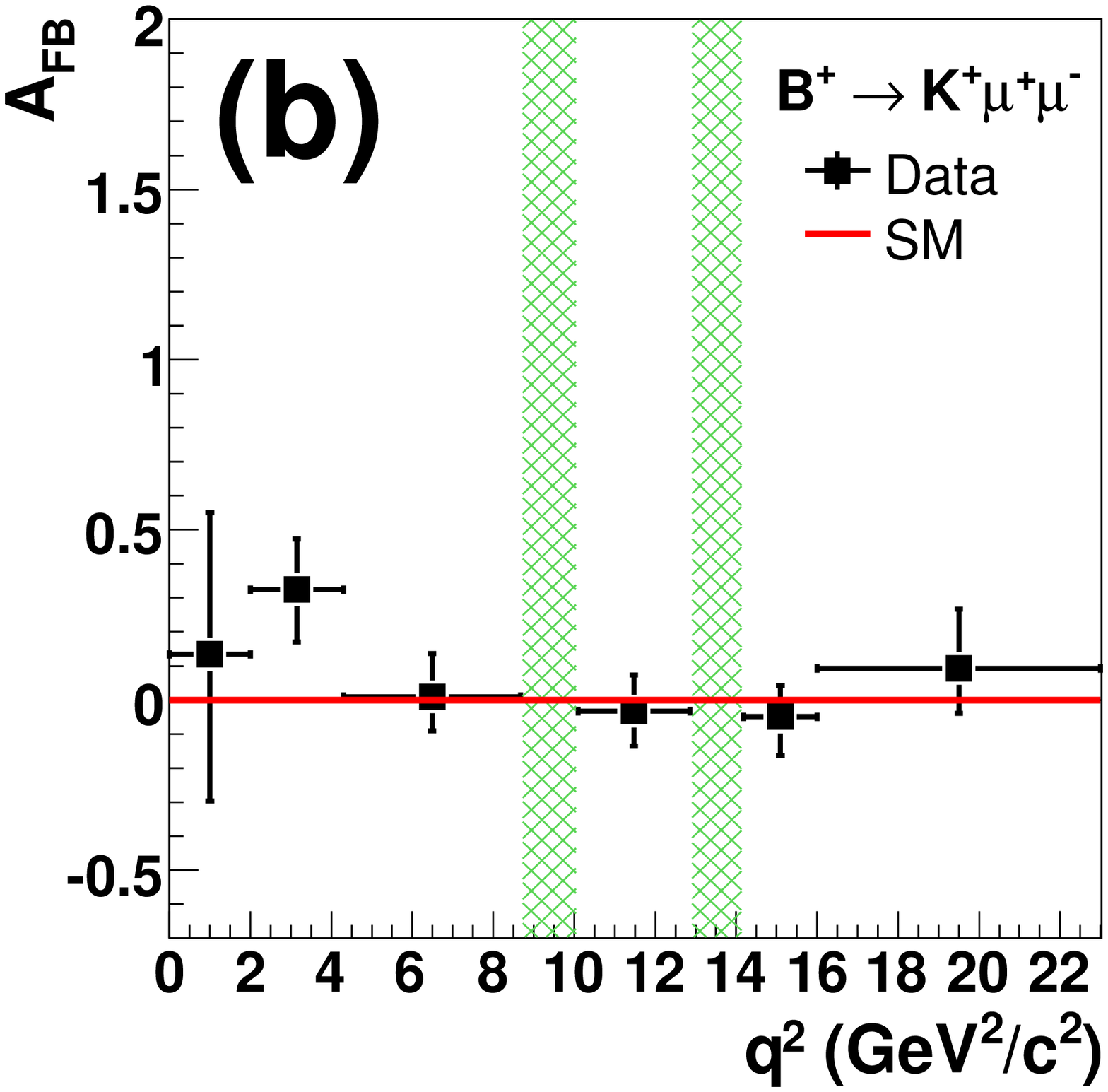}
    \end{tabular}
  \end{center}
  \caption{
Measurements of
forward-backward asymmetry $\afb$
in the decay (a) $\bzkstmm$ and
(b) $\bpkmm$
as a function of dimuon mass 
squared 
$q^2$.
 Points are the fit results from data.
The solid curves are the SM expectation~\cite{note:eos_curve}.
The dotted curve is the $C_7=-C_7^{\rm SM}$ expectation suggested by some BSM models.
Hatched regions are excluded resonant (charmonium) decay regions.}
  \label{fig:afb}
\end{figure}

\begin{figure}[htpb]
  \begin{center}
\setlength{\tabcolsep}{1pt}
\begin{tabular}{c}
\includegraphics[scale=0.245,trim=8 5 57 48,clip]{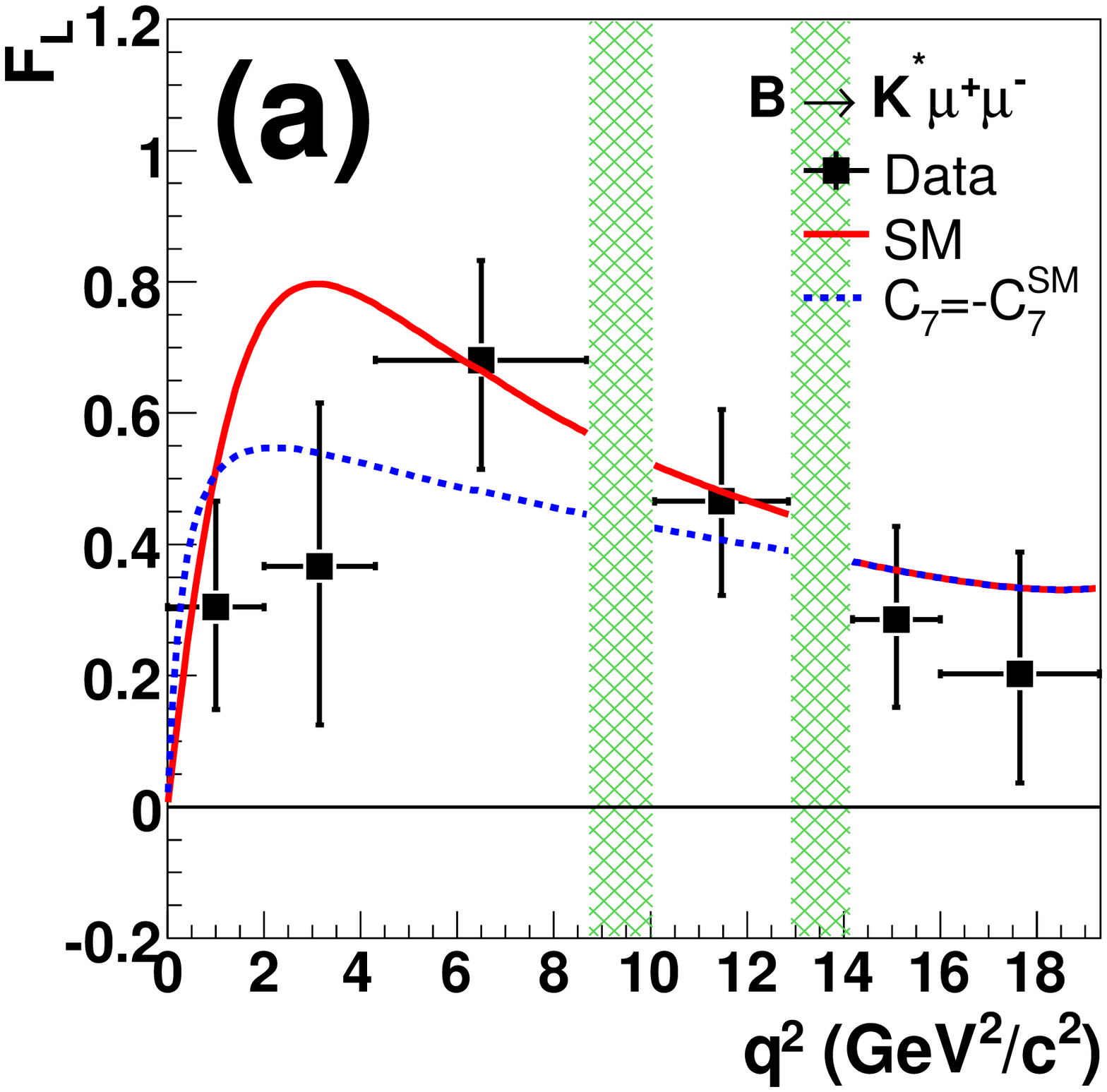}
\includegraphics[scale=0.245,trim=8 5 57 48,clip]{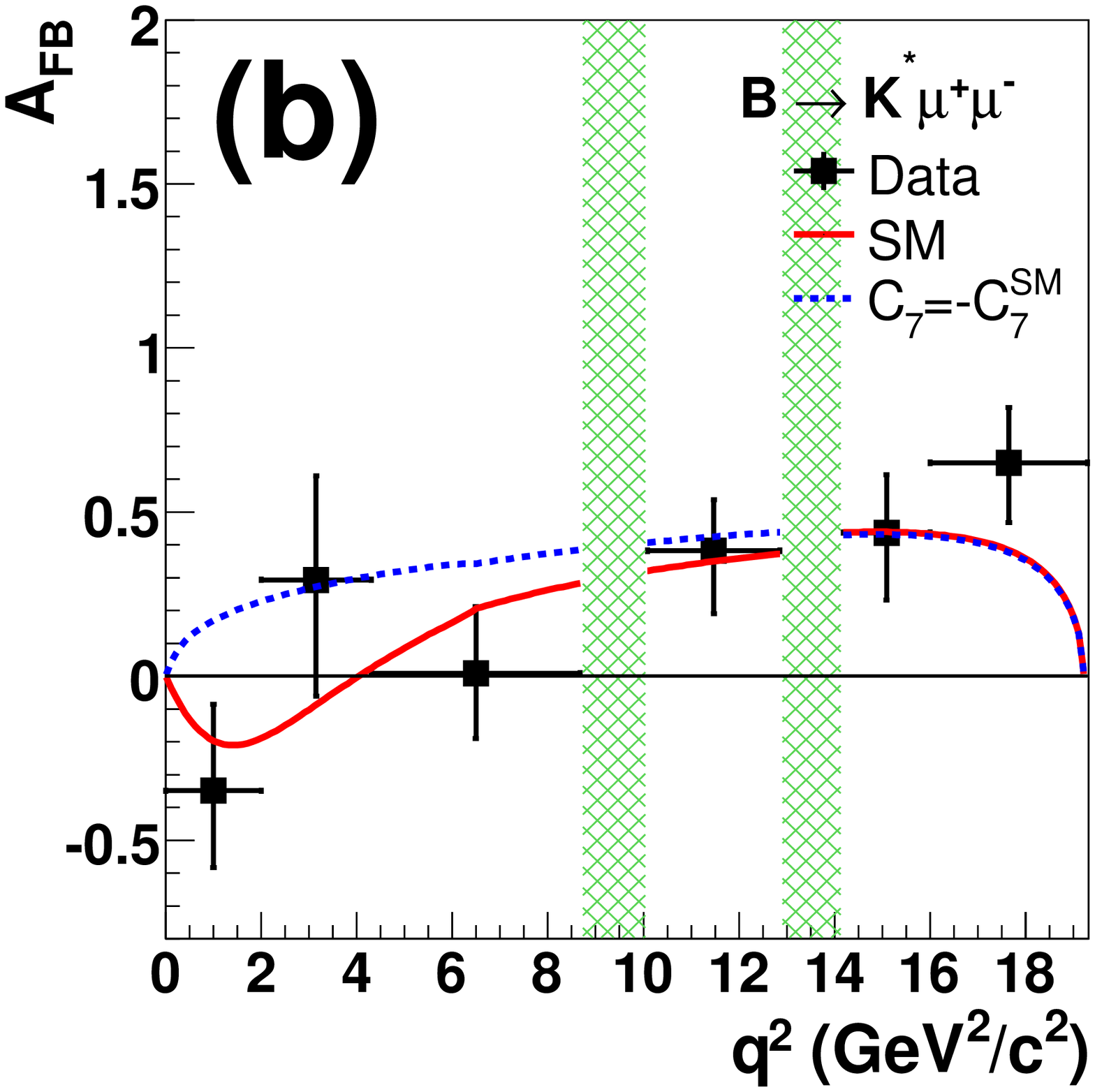} \\
\includegraphics[scale=0.245,trim=8 5 57 48,clip]{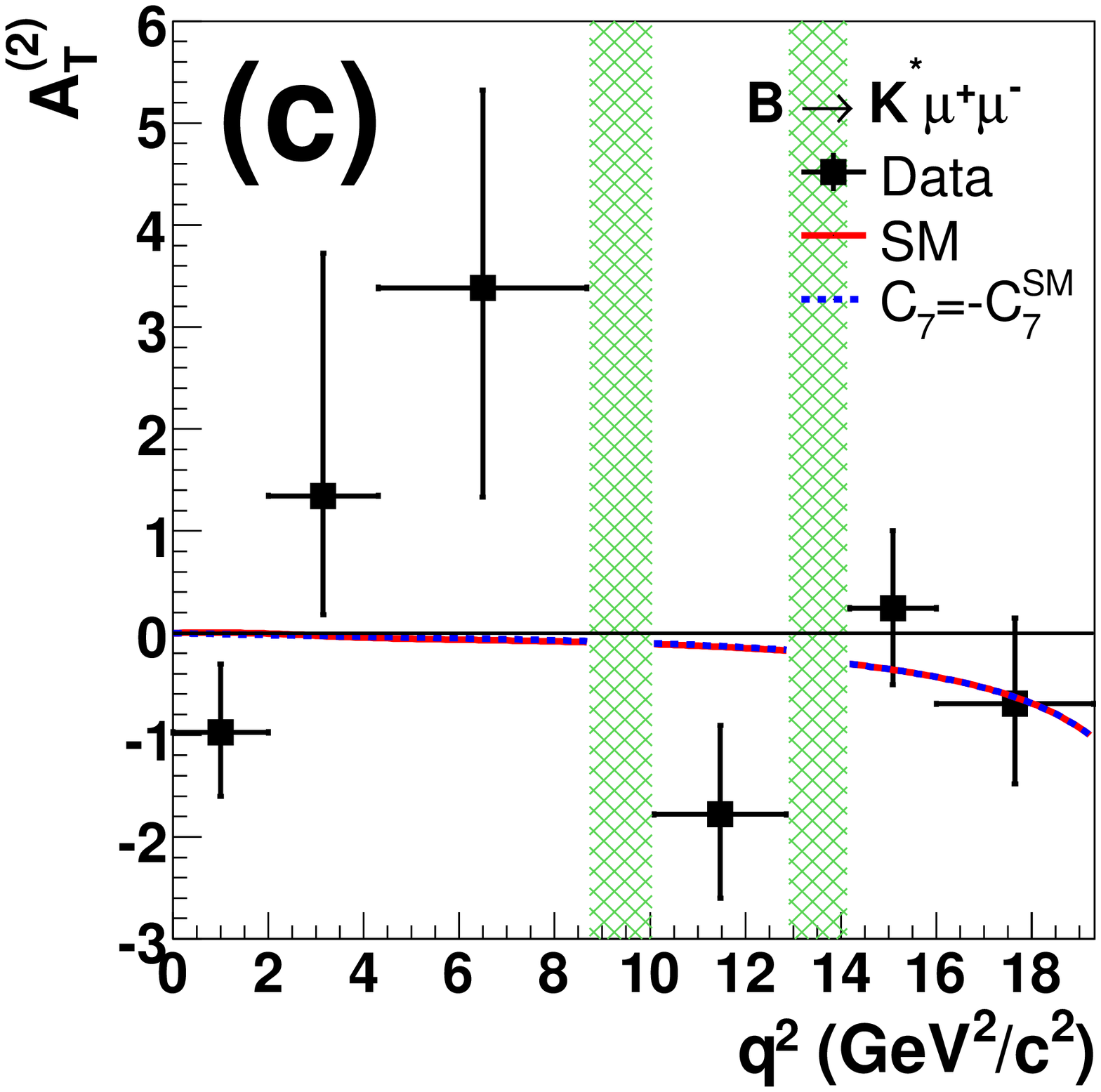}
\includegraphics[scale=0.245,trim=8 5 57 48,clip]{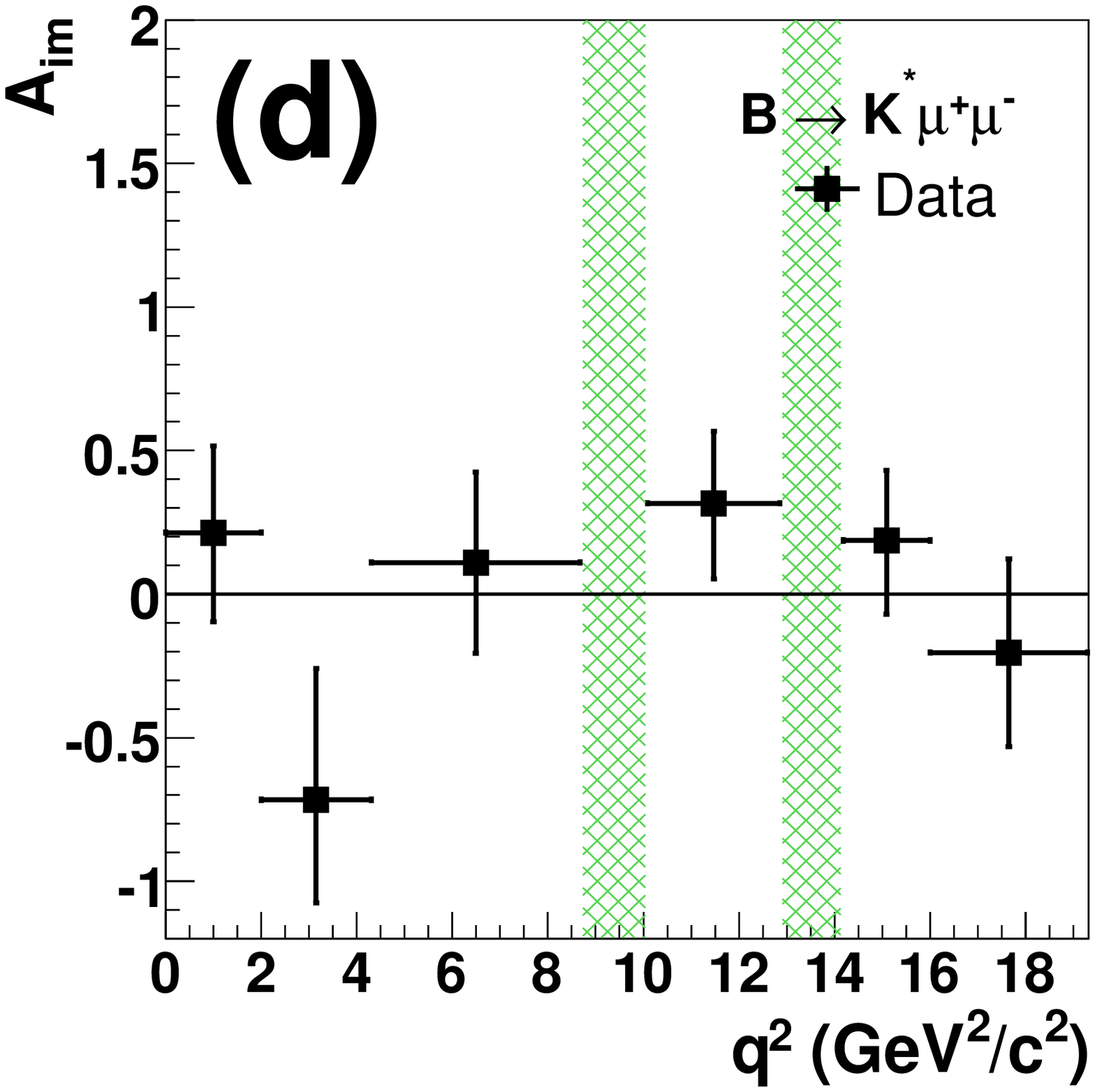} \\
    \end{tabular}
  \end{center}
  \caption{
Measurements of
(a) longitudinal $\kstall$ polarization fraction $\fl$,
(b) forward-backward asymmetry $\afb$,
(c) transverse polarization asymmetry $\attwo$,
and (d) $T$-odd $CP$ asymmetry $\aim$
in the combined decay mode $\bkstallmm$,
all presented
as a function of dimuon mass squared $q^2$.
The points are the fit results from data.
The solid curves are the SM expectation~\cite{note:eos_curve}.
The dotted curves are the $C_7=-C_7^{\rm SM}$ expectation.
Hatched regions are resonant (charmonium) decay regions.}

  \label{fig:fl_afb_at2_aim}
\end{figure}

The extracted observables in the decay mode $\bzkstmm$
are listed in Table~\ref{tab:summary_kstmm}, 
and the forward-backward asymmetry $\afb$ is illustrated 
in Fig.~\ref{fig:afb}(a) as a function of $q^2$. 
To increase sensitivity, we also perform a fit to the combined
$\bzkstmm$  and $\bpkstpmm$ modes,
assuming they have the same decay dynamics.
The combined fit results are listed in
Table~\ref{tab:summary_kstmm_all}  and shown in Fig.~\ref{fig:fl_afb_at2_aim}. 
On average $\afb$ resolutions are improved by factors of 1.5 (1.8) times from 
previous CDF measurement~\cite{Aaltonen:2011cn} for $\bzkstmm$ ($\bkstallmm$).
The current data are consistent with the SM and with the inverted $C_7$ scenario.
In addition, a new measurement of $\afb$
in the decay  $\bpkmm$ is obtained.
In the SM, the expected value of $\afb$ for this mode is quite small over the entire range of $q^2$~\cite{Bobeth:2007dw_APS},
whereas some BSM models predict enhanced values of $\afb$~\cite{Alok:2010zd_APS}.
In the $\afb$ fit, we assume no scalar term~\cite{Bobeth:2007dw_APS}
and  set $\fl=1$. 
The result is shown in Fig.~\ref{fig:afb}(b) and listed in  Table~\ref{tab:summary_kmm}.
The current data is consistent with the SM.

The sources of systematic uncertainty in the angular observables include the estimation of detector acceptance of the decay angles, 
signal fraction estimation and shape modeling of events in the signal window,
feed-down background from other $B$ decays,
trigger efficiency and bias modeling, incorrect $K$-$\pi$ assignment in the $\kst \to K^+ \pi^-$ decay,
and fitting bias. The largest contribution is that from 
uncertainties on the signal fraction in the signal window.
The total systematic uncertainties on the asymmetry observables for various $q^2$ values are in the ranges 
0.02 -- 0.09 for $\fl$, 0.05 -- 0.17 for $\afb$, 0.07 -- 2.21 for $\attwo$, 
and  0.01 -- 0.14 for $\aim$ in the mode $\bzkstmm$.
Similar sizes of systematic uncertainties are found in the combined fit results. 
 

\begin{table*}[t]
  \caption{Summary of $\bzkstmm$ fit results. First (second) uncertainty is statistical (systematic).}
  \begin{center}
    \begin{tabular}{ccrrrr}
      \hline
      \hline
      $q^2$ ($\GeV^2/c^2$) & $N(\kstmm)$ & \multicolumn{1}{c}{$\fl$} & \multicolumn{1}{c}{$\afb$} & \multicolumn{1}{c}{$\attwo$} & \multicolumn{1}{c}{$\aim$} \\
      \hline
      $[0.00,2.00)$   & $30.7\pm4.7$ & $0.31^{+0.17}_{-0.16}\pm 0.02$ & $-0.37^{+0.27}_{-0.32}\pm 0.11$ & $-0.8\pm0.7\pm 0.3$ & $0.37^{+0.31}_{-0.33}\pm 0.08$ \\
      $[2.00,4.30)$   & $14.2\pm4.2$ & $0.35^{+0.26}_{-0.24}\pm 0.03$ & $0.30^{+0.32}_{-0.36}\pm 0.17$ & $1.4^{+2.0}_{-1.1}\pm 1.2$ & $-0.80^{+0.48}_{-0.29}\pm 0.13$ \\
      $[4.30,8.68)$   & $31.3\pm7.4$ & $0.60^{+0.17}_{-0.18}\pm 0.05$ & $-0.08^{+0.22}_{-0.21}\pm 0.03$ & $1.8^{+1.6}_{-1.7}\pm 1.5$ & $0.03^{+0.34}_{-0.34}\pm 0.06$ \\
      $[10.09,12.86)$ & $38.4\pm7.6$ & $0.40^{+0.16}_{-0.16}\pm 0.02$ & $0.42^{+0.17}_{-0.21}\pm 0.10$ & $-1.0^{+0.9}_{-0.8}\pm 0.5$ & $0.47^{+0.26}_{-0.28}\pm 0.09$ \\
      $[14.18,16.00)$ & $31.6\pm4.7$ & $0.32^{+0.14}_{-0.14}\pm 0.03$ & $0.40^{+0.18}_{-0.21}\pm 0.07$ & $0.4\pm0.8\pm 0.2$ & $0.15^{+0.25}_{-0.26}\pm 0.01$ \\
      $[16.00,19.30)$ & $20.7\pm4.8$ & $0.16^{+0.22}_{-0.18}\pm 0.06$ & $0.66^{+0.18}_{-0.26}\pm 0.19$ & $-0.9\pm0.8\pm 0.4$ & $-0.30^{+0.36}_{-0.35}\pm 0.14$ \\
      \hline
      $[0.00,4.30)$   & $44.2\pm6.5$ & $0.33^{+0.14}_{-0.14}\pm 0.02$ & $-0.08^{+0.21}_{-0.20}\pm 0.05$ & $-0.2\pm0.6\pm 0.1$ & $-0.02^{+0.28}_{-0.28}\pm 0.01$ \\
      \hline
      $[1.00,6.00)$   & $23.8\pm6.5$ & $0.60^{+0.21}_{-0.23}\pm 0.09$ & $0.36^{+0.46}_{-0.28}\pm 0.11$ & $1.6^{+1.8}_{-1.9}\pm 2.2$ & $-0.02^{+0.40}_{-0.40}\pm 0.03$ \\
      \hline
      \hline
    \end{tabular}
  \end{center}
  \label{tab:summary_kstmm}
\end{table*}

\begin{table*}[htpb]

  \caption{Summary of combined $\bkstallmm$ fit results. First (second) uncertainty is statistical (systematic).
$N(\kstmm)$ is taken from Table~\ref{tab:summary_kstmm}.}
  \begin{center}
    \begin{tabular}{ccrrrr}
      \hline
      \hline
      $q^2$ ($\GeV^2/c^2$) & $N(\kstpmm)$ & \multicolumn{1}{c}{$\fl$} & \multicolumn{1}{c}{$\afb$} & \multicolumn{1}{c}{$\attwo$} & \multicolumn{1}{c}{$\aim$} \\
      \hline
      $[0.00,2.00)$   & $2.5\pm1.6$ & $0.30^{+0.16}_{-0.16}\pm 0.02$ & $-0.35^{+0.26}_{-0.23}\pm 0.10$ & $-1.0^{+0.7}_{-0.6}\pm 0.4$ & $0.21^{+0.30}_{-0.31}\pm 0.10$ \\
      $[2.00,4.30)$   & $1.3\pm1.8$ & $0.37^{+0.25}_{-0.24}\pm 0.10$ & $0.29^{+0.32}_{-0.35}\pm 0.15$ & $1.3^{+2.4}_{-1.2}\pm 0.9$ & $-0.72^{+0.46}_{-0.36}\pm 0.21$ \\
      $[4.30,8.68)$   & $3.9\pm3.5$ & $0.68^{+0.15}_{-0.17}\pm 0.09$ & $0.01^{+0.20}_{-0.20}\pm 0.09$ & $3.4^{+1.9}_{-2.1}\pm 3.6$ & $0.11^{+0.31}_{-0.32}\pm 0.09$ \\
      $[10.09,12.86)$ & $6.0\pm2.8$ & $0.47^{+0.14}_{-0.14}\pm 0.03$ & $0.38^{+0.16}_{-0.19}\pm 0.09$ & $-1.8^{+0.9}_{-0.8}\pm 0.8$ & $0.32^{+0.25}_{-0.26}\pm 0.06$ \\
      $[14.18,16.00)$ & $1.6\pm1.8$ & $0.29^{+0.14}_{-0.13}\pm 0.05$ & $0.44^{+0.18}_{-0.21}\pm 0.10$ & $0.2\pm0.8\pm 0.2$ & $0.19^{+0.24}_{-0.26}\pm 0.04$ \\
      $[16.00,19.30)$ & $4.1\pm2.3$ & $0.20^{+0.19}_{-0.17}\pm 0.05$ & $0.65^{+0.17}_{-0.18}\pm 0.16$ & $-0.7\pm0.8\pm 0.3$ & $-0.20^{+0.33}_{-0.33}\pm 0.09$ \\
      \hline	      			     
      $[0.00,4.30)$   & $3.8\pm2.4$ & $0.33^{+0.14}_{-0.13}\pm 0.03$ & $-0.08^{+0.21}_{-0.20}\pm 0.05$ & $-0.3\pm0.6\pm 0.1$ & $-0.10^{+0.27}_{-0.26}\pm 0.10$ \\
      \hline	      			     
      $[1.00,6.00)$   & $5.0\pm3.0$ & $0.69^{+0.19}_{-0.21}\pm 0.08$ & $0.29^{+0.20}_{-0.23}\pm 0.07$ & $1.7\pm2.2\pm 2.5$ & $0.09^{+0.34}_{-0.35}\pm 0.18$ \\
      \hline
      \hline
    \end{tabular}
  \end{center}
  \label{tab:summary_kstmm_all}
\end{table*}

\begin{table}[htpb]
  \caption{Summary of $\bpkmm$ fit results.}
  \begin{center}
     \begin{tabular}{ccr}
      \hline
      \hline
      $q^2$ ($\GeV^2/c^2$)  & $N(\kmm)$ & \multicolumn{1}{c}{$\afb$} \\
      \hline
      $[0.00,2.00)$ & $18.6\pm5.6$ & $0.13^{+0.42}_{-0.43}\pm 0.07$ \\
      $[2.00,4.30)$ & $40.3\pm6.7$ & $0.32^{+0.15}_{-0.16}\pm 0.05$ \\
      $[4.30,8.68)$ & $68.5\pm10.5$ & $0.01^{+0.13}_{-0.10}\pm 0.01$ \\
      $[10.09,12.86)$ & $43.5\pm7.1$ & $-0.03^{+0.11}_{-0.10}\pm 0.04$ \\
      $[14.18,16.00)$ & $35.9\pm5.7$ & $-0.05^{+0.09}_{-0.11}\pm 0.03$ \\
      $[16.00,23.00)$ & $28.9\pm6.3$ & $0.09^{+0.17}_{-0.13}\pm 0.03$ \\
      \hline
      $[0.00,4.30)$ & $57.8\pm8.8$ & $0.31^{+0.16}_{-0.16}\pm 0.04$ \\
      \hline
      $[1.00,6.00)$ & $74.5\pm9.6$ & $0.13^{+0.09}_{-0.09}\pm 0.02$ \\
      \hline
      \hline
    \end{tabular}
  \end{center}
  \label{tab:summary_kmm}
\end{table}

In summary, 
we have reconstructed the decays
$\bzkstmm$ and $\bpkstpmm$ and measured their angular distributions. 
We have measured
the  muon forward-backward asymmetry $\afb$, 
the $\kstall$  longitudinal polarization fraction $\fl$, 
the transverse polarization asymmetry $\attwo$, and the 
$T$-odd $CP$ asymmetry $\aim$
as a function of the dimuon mass squared $q^2$.
Measurements of  $\attwo$ and $\aim$ are reported
for the first time.
The muon forward-backward asymmetry $\afb$ 
is also measured in the decay mode $\bpkmm$
and represents an update with higher precision.
All of the new reported results 
presented in this Letter are among the most 
accurate to date and consistent with the SM predictions,
but still statistically limited 
in providing stringent tests on various models.
The results are also consistent 
with recent measurements from $B$-factory experiments~\cite{Aubert:2008ju_APS,Wei:2009zv}.

\begin{acknowledgments}
We wish to tender our cordial thanks to Christoph Bobeth, Joaquim Matias, and Wolfgang Altmannshofer,
for close communication and valuable suggestions about the angular observables.
We express our special gratitude to Danny van Dyk, who provided the tool to obtain the theoretical predictions for the angular observables.

We thank the Fermilab staff and the technical staffs of the participating institutions for their vital contributions. This work was supported by the U.S. Department of Energy and National Science Foundation; the Italian Istituto Nazionale di Fisica Nucleare; the Ministry of Education, Culture, Sports, Science and Technology of Japan; the Natural Sciences and Engineering Research Council of Canada; the National Science Council of the Republic of China; the Swiss National Science Foundation; the A.P. Sloan Foundation; the Bundesministerium f\"ur Bildung und Forschung, Germany; the Korean World Class University Program, the National Research Foundation of Korea; the Science and Technology Facilities Council and the Royal Society, UK; the Institut National de Physique Nucleaire et Physique des Particules/CNRS; the Russian Foundation for Basic Research; the Ministerio de Ciencia e Innovaci\'{o}n, and Programa Consolider-Ingenio 2010, Spain; the Slovak R\&D Agency; and the Academy of Finland.
\end{acknowledgments}

\bibliographystyle{h-physrev5_cdf_prl}
\bibliography{reference}

\begin{thebibliography}{10}
\bibitem{CDF_coordinate}
We use a cylindrical coordinate system in which $\theta$ is the polar angle
  with respect to the proton beamline and pseudorapidity $\eta \equiv -\ln(\tan
  \theta/2)$.
\end{thebibliography}

\end{document}